\documentclass[prl,aps,twocolumn]{revtex4}
\usepackage{graphicx}
\usepackage{amsmath,amssymb} 
\def\strutdepth{\dp\strutbox}
\def\nw#1{\strut\vadjust{\kern-\strutdepth\vtop to0pt{\vss\hbox to\hsize
{\hskip\hsize\hskip5pt$\leftarrow$\hss\strut}}}{\em #1}}
\begin{document}

\title{Air Entrainment by Viscous Contact Lines}
\author{Antonin Marchand$^1$, Tak Shing Chan$^2$,  Jacco H. Snoeijer$^2$ and Bruno Andreotti$^1$}
\affiliation{
$^{1}$Physique et M\'ecanique des Milieux H\'et\'erog\`enes, UMR
7636 ESPCI -CNRS, Univ. Paris-Diderot, 10 rue Vauquelin, 75005, Paris.\\
$^{2}$Physics of Fluids Group, Faculty of Science and Technology and Mesa+ Institute, University of Twente, 7500AE Enschede, The Netherlands.
}
\date{\today}

\begin{abstract}
The entrainment of air by advancing contact lines is studied by plunging a solid plate into a very viscous liquid. Above a threshold velocity, we observe the formation of an extended air film, typically $10$ microns thick, which subsequently decays into air bubbles. Exploring a large range of viscous liquids, we find an unexpectedly weak dependence of entrainment speed on liquid viscosity, pointing towards a crucial role of the flow inside the air film. This induces a striking asymmetry between wetting and dewetting: while the breakup of the air film strongly resembles the dewetting of a liquid film, the wetting speeds are larger by orders of magnitude.
\end{abstract}

\maketitle

Objects that impact on a liquid interface can entrain small bubbles of air into the liquid. This happens for example when raindrops fall in the ocean~\cite{oguz95} or when liquid is poured into a reservoir at sufficiently large speeds~\cite{E01,LQE04}. Such entrainment of air is often a limiting factor in industrial applications such as coating and nano-scale printing techniques, where the bubbles disturb the process~\cite{BR79,JJB06}. A well studied case is the entrainment of air by very viscous jets impacting on a reservoir of the same liquid~\cite{JNRR91,JM92,E01,LQE04}. The onset of entrainment  is essentially determined by the properties of the liquid, $U_e \sim \gamma/\eta_\ell$, which reflects a balance of the liquid viscosity $\eta_\ell$ and the surface tension $\gamma$. Changing the nature of the gas only has a minor influence on the entrainment process~\cite{E01,LQE04}.

A very different picture has emerged recently in the context of drops impacting on a wall, for which the presence of air has a dramatic effect~~\cite{XZN05,Dri10,TVRL10}. It was found that splashing can be suppressed completely by reducing the air pressure to about a third of the atmospheric pressure. This caused huge excitement, in particular because such a pressure reduction does not lead to any change of the gas viscosity $\eta_g$~\cite{JRZ10,Mandre09,MMB10}: pressure only affects the gas density, and thus the speed of sound and the mean free path in the gas. A similar paradox is encountered for air entrainment by rapidly advancing contact lines, where a liquid advances over a surface that it partially wets~\cite{DYCB07,LRHP11,BR79,BK76b,SK00,Ben07}. Once again, it was found that depressurizing the gas leads to a significant increase of the threshold of air entrainment~\cite{Ben07,Ben10}. This contradicts the classical viewpoint that, for given wettability, the contact line speed depends mainly on the liquid properties as $ \sim \gamma/\eta_\ell$~\cite{C86,SDAF06,DYCB07,BEIMR09}, with minor influence of the gaseous phase.

\begin{figure}[t!]
\includegraphics{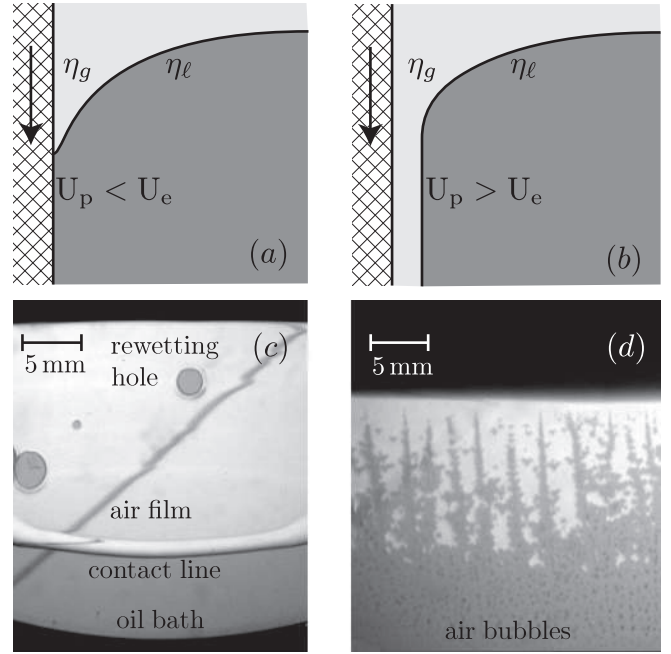}
\caption{Air entrainment by a contact line when a solid plate is plunged into a viscous liquid. (a,b) Sketch of dynamical meniscus below the threshold of air entrainment (a), and above the threshold of air entrainment (b). In the latter case a film of thickness $h$ is formed. (c,d) Experimental images of the entrained air film in silicone oils of varying viscosity $\eta_\ell$. (c) For $\eta_\ell= 0.1\, {\rm Pa.s}$, an extended air film is entrained behind the contact line moving downwards into the bath. The film is destroyed by the formation of ``rewetting" holes. The diagonal dark line is the reflection of a straight wire that gives an impression of the interface profile. (d) For a more viscous oil, $\eta_\ell = 5 \,{\rm Pa.s}$, the entrained air film is rapidly destabilized to form small bubbles (dark spots).}
\label{fig.sketch}
\vspace{-1mm}
\end{figure}

In this Letter we reveal the role of the air for advancing contact lines in a paradigmatic system: a partially wetting solid plate is plunged into a reservoir of viscous liquid (Fig.~\ref{fig.sketch}). At sufficiently high speeds we observe the entrainment of an air film, typically $10\,$microns thickness, which subsequently decays into bubbles. The liquid viscosity $\eta_\ell$ is varied over more than two decades by using silicon oils of different molecular weights. It is found that the entrainment speed $U_e$ changes much less than the expected scaling $\sim 1/\eta_\ell$. Using an approximate hydrodynamic model we argue that this can be attributed to the flow of air into the strongly confined film, making the contact line velocity strongly dependent on both gas and liquid viscosities. This induces a striking asymmetry between wetting and dewetting: the same liquid can advance much faster than it recedes, by orders of magnitude.

The experiments are carried using silicon oils (PDMS, Rhodorsil 47V series) with dynamic viscosities $\eta_\ell= 0.02, 0.10,0.5,1.0$  and $5~$Pa.s. These liquids are essentially non-volatile, insensitive to contamination, while the surface tension $\gamma=22~\mathrm{mN.m^{-1}}$ and density $\rho=980~\mathrm{kg.m^{-3}}$ are approximately constant for all viscosities. The reservoir is a transparent acrylic container of size is $29$x$15$x$13.5\,$cm, which is much larger than the capillary length $\ell_{\gamma}=(\gamma/\rho g)^{1/2} =1.5\,$~mm. The substrate consists of a silicon wafer (circular, diameter $10\,$cm), which is coated by a thin layer of fluorinated material (FC 725 (3M) in ethyl acetate). For all liquids, this results into static contact angles between $51^\circ$ and $57^\circ$. The wafer is clamped onto a $10$~mm thick metallic blade screwed to a $50$~cm long high-speed linear stage. The combination of using controlled speeds and very viscous liquids  one avoids complexities of splashing as well as the formation of interface cusps~\cite{BR79,BK76b,DYCB07,SK00,Ben07}. In addition the effect of inertia is eliminated both in the gas and in the liquid; the relevant Reynolds numbers are at worst $\sim 1$, but typically orders of magnitude smaller. For each liquid, we plunge the wafer into the reservoir at different plate velocities $U_p$, up to $0.7\,$m/s. The process is recorded using a high-speed Photron SA3 camera ($2000\,$Hz, $1024$x$1024\,$pixels). The plate and contact line velocities are then extracted from space-time diagrams using a correlation technique with a sub-pixel resolution, leading to a precision better than the percent. Reproductibility is achieved within $15\%$. The film thickness $h$ is determined from an accurate measurement of the volume of air entrained in the bath. The air bubbles trapped at the end of an experiment are imaged with Nikon D300s ($4288$x$2848$~pix) mounted with macro-lens. The size of each bubble is determined automatically within a $10~\mu$m resolution, by fitting an elliptic shape. Most of the uncertainty on $h$ results from the estimate of the surface covered by the air film.

The experimental scenario is presented in Fig.~\ref{fig.sketch}. At small speeds we observe that the contact line equilibrates to form a stationary meniscus and no air is entrained into the liquid (Fig.~\ref{fig.sketch}a). Above a critical velocity, however, the contact line keeps moving downward into the reservoir and deposits a thin film of air (Fig.~\ref{fig.sketch}b). Subsequently, the air film breaks up into small air bubbles. The dynamical structures that appear during the breakup of the film turn out to depend strongly on the oil viscosity, as can be seen from the experimental images of Fig.~\ref{fig.sketch}cd. For the most viscous liquids the air films are very fragile and rapidly break up into smaller bubbles (Fig.~\ref{fig.sketch}d). However, for the lower viscosities one observes the formation of very extended air films (Fig.~\ref{fig.sketch}c). At the front of the film, the contact line develops a ridge-like structure that is common for dewetting of liquid films~\cite{RBR91,FK04,SnE10}. The peculiarity of the present experiment is that in this case the \emph{air} is dewetted, not the liquid.

\begin{figure}[t!]
\includegraphics{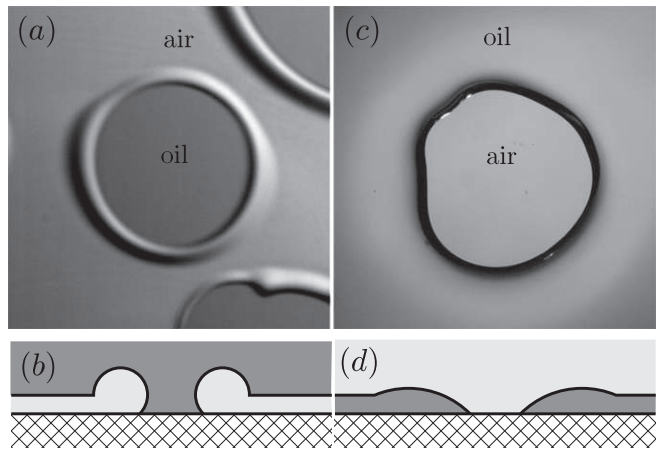}
\caption{``Rewetting" holes (air film invaded by liquid) as the inverse of the classical dewetting of a liquid film. (a,b) Rewetting: once the silicon oil reestablishes contact with the solid, one observes the growth of a circular zone that invades the air film. The moving front collects the air in a thick rim that is clearly visible in the image. Typical contact line speeds range from $1$ to $10\,$cm/s for the silicone oils used in this study. (c,d) Dewetting: a thin film of silicon oil dewets in the form a circular hole. Note that for the same liquids as in (a) the typical dewetting speeds are smaller by orders of magnitude (from $40\,\mu$m/s to $1\,$cm/s).
}
\label{fig.rim}
\vspace{-1mm}
\end{figure}

An even more striking analogy with classical dewetting of liquids is the nucleation of nearly circular regions inside the film (Fig.~\ref{fig.sketch}c). However, the circles now represent regions of \emph{rewetting}, where the liquid reestablishes the contact with the solid. These ``rewetting holes" can be considered as the inverse of the ``dewetting holes", since the roles of air and liquid are exchanged. Figure~\ref{fig.rim}a shows a close-up of a rewetting hole (cross-section sketched in Fig.~\ref{fig.rim}b). The radius of the holes increases linearly with time, and the advancing contact lines collects the air inside a thick rim. While this is analogous to the inverse problem of the dewetting holes (Fig.~\ref{fig.rim}cd), the process is by no means symmetric: the rewetting holes grow with a velocity $U_e$ that is orders of magnitude faster than their dewetting counterparts, up to a factor $1000$ for the liquids used in this study.
\begin{figure}[t!]
\includegraphics{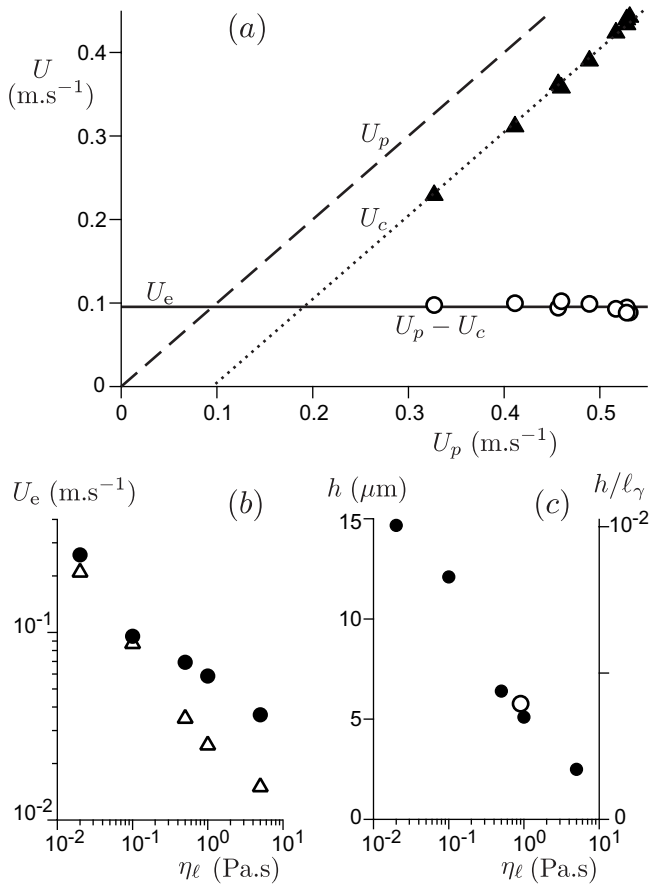}
\caption{(a) Contact line velocity $U_{c}$ (triangles) as a function of the plate velocity $U_p$, for $\eta_\ell$=0.1Pa.s. The relative velocity $U_p - U_c$ (circles) is independent of the plate speed.
(b) Entrainment speed $U_e$ for different liquid viscosities $\eta_\ell$, measured in two ways: the circles represent the relative velocity with respect to the plate, $U_p-U_{cl}$, while the triangles are the rewetting speeds of the holes as in Fig.~\ref{fig.rim}. (c) Film thickness $h$ for different $\eta_\ell$, taken at constant plate velocity $U_p=0.67~\mathrm{m.s^{-1}}$. The open circle was taken for a liquid jet of glycerol entraining air at the same speed~\cite{LQE04}.
}
\label{fig.speed}
\vspace{-1mm}
\end{figure}

We further quantify the velocity of air entrainment for different liquid viscosities. A first measurement of $U_e$ is obtained from the growth velocities of the rewetting holes as in Fig.~\ref{fig.rim}. A second entrainment velocity can be extracted by a selecting a central part of the front of the air film, for which we obtain the contact line velocity $U_{cl}$ in the frame of the liquid reservoir. Figure~\ref{fig.speed}a reports the measured values for $\eta_\ell=0.1$~Pa.s, showing that the contact line velocity increases linearly with plate velocity $U_p$. Interestingly, however, the relative velocity, $U_{p}-U_{cl}$, turns out to be independent of the plate velocity and therefore seems to be an intrinsic property of the advancing contact line. This is why we may consider the growth velocity of the rewetting holes to be an independent measurement of the entrainment velocity. The resulting entrainment velocities $U_e$ are shown in Fig.~\ref{fig.speed}b, as a function of liquid viscosity $\eta_\ell$. Indeed, the two experimental definitions of $U_e$ agree very well for the smallest $\eta_\ell$ (closed circles are based on the front of the film, open triangles correspond to rewetting holes). For larger $\eta_\ell$, the film rapidly destabilizes and it is more difficult to define the front of the film. This induces a difference between the two types of velocity measurements of about a factor 2; the hole velocities are certainly more reproducible in this very viscous regime.
\begin{figure}[t!]
\includegraphics{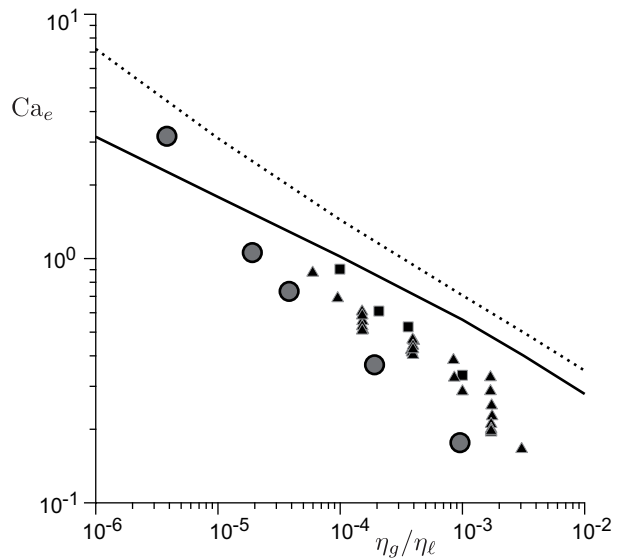}
\caption{Dimensionless entrainment speed, ${\rm Ca} = U_e \eta_\ell/\gamma$, versus the viscosity ratio $\eta_g/\eta_\ell$ for: silicon-oil/air ($\bullet$), silicon-oil/air ($\blacksquare$ after \cite{Ben07}) and various liquids/air ($\blacktriangle$ after \cite{BK76a}). Oil slip lengths $\lambda_\ell=10^{-5} \ell_\gamma$, and air slip lengths $\lambda_g=10^{-4}\ell_\gamma$ (solid line) and $10^{-2}\ell_\gamma$ (dotted line), which correspond to mean free paths $\ell_{MF}=70$ nm (solid) and 7 $\mu$m.
}
\label{fig.dimensionless}
\vspace{-1mm}
\end{figure}

The key result of the velocity measurements is that, although $U_e$ decreases with liquid viscosity, the dependence is clearly much weaker than the expected $\sim 1/\eta_\ell$. The entrainment speed is reduced by a factor $10$, while viscosity is varied by a factor 250 (Fig.~\ref{fig.speed}b). Since the liquid inertia is negligible for these highly viscous liquids, this means that the properties of the air must have a significant influence on the entrainment speed. On the other hand, the speed is not determined by the air alone, since that would yield no dependence on $\eta_\ell$ at all. To reveal the interplay between air and liquid phases, we introduce a dimensionless capillary number, ${\rm Ca}=U_e \eta_\ell / \gamma$, that is based on the liquid viscosity. The experimental results are represented in Fig.~\ref{fig.dimensionless}, showing ${\rm Ca}$ versus the ratio of gas and liquid viscosities $\eta_g/\eta_\ell$ (closed circles). Clearly, the capillary number for air entrainment displays a dependence that is much stronger than $\sim \ln(\eta_\ell/\eta_g)$, which is the scaling for air entrainment by liquid jets~\cite{E01,LQE04} and the prediction by Ref.~\cite{C86}. The air thus has a much larger influence than expected. On the same figure we collected data from the coating literature, based on tapes running continuously in a bath,  showing a similar trend (various symbols, see caption). Note that in these experiments the contact line typically develops a sharp cusp from which small air bubbles are emitted, rather than an extended air film.
\begin{figure}[t!]
\includegraphics{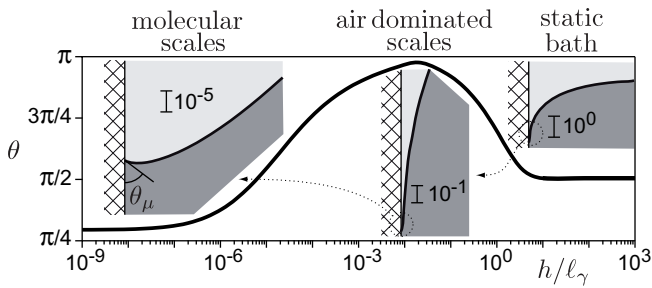}
\caption{Insets: shape of the dynamical meniscus at the entrainment threshold for $\eta_g/\eta_\ell=10^{-4}$, as predicted by the hydrodynamic model~\cite{sup}. Main graph: local slope $\theta$ as a function of the local thickness $h$. Lengths are in units of $\ell_\gamma$.}
\label{fig.model}
\vspace{-1mm}
\end{figure}

In order to understand the influence of air, one can extend the common lubrication approximation to large slopes and 2 phase flow, but small curvature of the interface.  In the spirit of~\cite{SN06,Hew09}, one can start from the analytical solutions of flow in a wedge due to Huh \& Scriven~\cite{HS71}, as detailed in the Supplementary Material~\cite{sup}. Numerical solution of the model provides the provides the capillary number for entrainment ${\rm Ca}_e$ shown in Fig.~\ref{fig.dimensionless} as the solid line. This captures the order of magnitude for the entrainment speed as well as the unexpectedly strong dependence on viscosity ratio. However, due to the strong curvature of the interface, the approximation cannot be expected to give a fully quantitative prediction. Within this model, the viscous dissipation rates, $\dot{E}_g$ and $\dot{E}_\ell$ in the gas and the liquid can be estimated as a function of the local slope $\theta$:
\begin{eqnarray}
\dot{E}_g \sim \frac{\eta_gU^2}{\pi-\theta}, \quad \dot{E}_\ell \sim \eta_\ell U^2 (\pi-\theta)^2,
\end{eqnarray}
which is slightly different from the estimates in~\cite{DYCB07,LRHP11}. Clearly, the dissipation in the gas can dominate over liquid dissipation for contact angles very close to $\pi$, with a cross-over occurring when $\eta_g /\eta_\ell \sim (\pi -\theta)^3$. As shown in Fig.~\ref{fig.model}, such angles indeed appear naturally when the liquid advances rapidly,  while this is not the case when the liquid recedes. This explains the marked asymmetry between wetting and dewetting: the faster advancing speed is due to the reduction of dissipation near $\theta \approx \pi$.

In this Letter we experimentally showed that the entrainment speeds of advancing contact lines do not scale as $\gamma/\eta_\ell$, but has a much weaker variation with liquid viscosity. We explain this by the influence of the air flow when the local angle of the interface is close to $\pi$. Can such a scenario explain the observed increase of entrainment speed when depressurizing the air? A pressure reduction does not affect the gas viscosity, but it does increase the mean free path by a factor $\sim p_{\rm atm}/p$. Since the mean free path under ambient conditions is approximately $70$~nm, it is pushed well into the micron range when pressure is reduced by a factor $100$ \cite{Ben07,Ben10}. The mean free path then becomes  comparable to the film thickness measured experimentally (Fig.~\ref{fig.speed}c). Interpreting the mean free path of the gas as an effective slip~\cite{sup,Boc93}, this indeed yields an increase of ${\rm Ca}_e$ (dotted line in Fig.~\ref{fig.dimensionless}). This suggests that depressurized air acts as a Knudsen gas when confined close to the contact line.

\begin{acknowledgments}

\textbf{Acknowledgments~--~} We are grateful to J. Eggers and K. Winkels for valuable discussions. M. Fruchart is thanked for his help during preliminary experiments. TSC acknowledges financial support by the FP7 Marie Curie Initial Training Network ``Surface Physics for Advanced Manufacturing'' project ITN 215723.
\end{acknowledgments}

\bibliography{all_ref}

\end{document}